# THE LEXINGTON BENCHMARKS FOR NUMERICAL SIMULATIONS OF NEBULAE

G. Ferland[1], L. Binette[2], M. Contini[3], J. Harrington[4], T. Kallman[5], H. Netzer[3], D. Péquignot[6], J. Raymond[7], R. Rubin[8], G. Shields[9], R. Sutherland[10], S. Viegas[11]

## Abstract

We present the results of a meeting on numerical simulations of ionized nebulae held at the University of Kentucky in conjunction with the celebration of the 70th birthdays of Profs. Donald Osterbrock and Michael Seaton.

## 1. Introduction

Numerical simulations of emission line regions, whether photo or shock ionized, are a vital tool in the analysis and interpretation of spectroscopic observations. Models can determine characteristics of the central source of ionizing radiation, the composition and conditions within the emitting gas, or, for shocks, the shock velocity. Osterbrock (1989) and Draine and McKee (1993) review the basic physical processes in these environments.

Although numerical simulations are a powerful tool, this capability is somewhat mitigated by the complexity of the calculations. There will always be underlying questions regarding the astronomical environment (i.e., the shape of the ionizing continuum, inhomogeneities, or the composition of the gas) and uncertainties introduced by the evolving atomic/molecular data base. On top of this, however, the numerical approximations, assumptions, and the complexity of the simulations themselves introduce an uncertainty that cannot be judged from a single calculation.

[1] Physics and Astronomy, University of Kentucky, Lexington, KY 40506, gary@cloud9.pa.uky.edu

[2] ESO, D-85748, Garching bei Muenchen, Germany, lbinette@eso.org

[3] School of Physics and Astronomy, Tel Aviv University, 69978 Tel Aviv, Israel, netzer@wise.tau.ac.il, contini@ccsg.tau.ac.il

[4] Astronomy, U of Maryland, College Park, MD 20742, jph@astro.umd.edu

[5] Code 665, NASA Goddard SFC, Greenbelt, MD 20771, tim@xstar.gsfc.nasa.gov

[6] Observatoire de Paris, Meudon F-92195, Meudon Principal Cedex, France, pequignot@obspm.fr

[7] CfA, 60 Garden St., Cambridge, MA 02138, raymond@cfassp8.harvard.edu

[8] NASA/Ames Research Center, MS 245-6, Moffett Field, CA 94035-1000, rubin@cygnus.arc.nasa.gov

[9] Astronomy, University of Texas, Austin, TX 78712, shields@astro.as.utexas.edu

[10] JILA, University of Colorado, Boulder, CO 80309-0440, ralph@zwicky.colorado.edu

[11] IAGUSP, Av. Miguel Stefano 4200, 04301 São Paulo, S.P., Brazil, viegas@iag.usp.ansp.br

With these questions in mind Daniel Péquignot held a meeting on model nebulae in Meudon, France, in 1985. This provided a forum where investigators could carefully compare model predictions and identify methods, assumptions, or atomic data which led to significant differences in results. The results of this meeting were a set of benchmarks which represented the state of the art at that time (Péquignot 1986). A second meeting on model nebulae was held in Lexington in association with the celebration of the seventieth birthdays of Professors Osterbrock and Seaton, who both defined the field and took an active part in the deliberations. This is a summary of the results of that workshop.

## 2. The Codes

Brief summaries of the codes follow: the lead author(s) of the code is followed by the identification number used in the following tables.

**Binette**: (1) MAPPINGS I is a combination of a shock code developed by Mike Dopita and a photoionization code developed by Luc Binette. Work started in 1980 and a first version is described in Binette et al. (1985). Outward-only continuum transfer is assumed. Recent revisions include a 5-level hydrogen atom which allows calculations at much higher densities and the effects of dust scattering and absorption on the emergent line spectrum. A description of the dust transfer and of most recent modifications can be found in Binette et al. (1993a, b). Instructions for obtaining the code via anonymous ftp may be obtained from the author.

Table 1 Parameters for Photoionized Nebulae

| Parameter\Table | 2 | 3 | 4 | 5 | 6 | 7 | 8 |
|---|---|---|---|---|---|---|---|
| $Q(H)$ or $\phi(H)$ | 1(49) | 4.26(49) | 1(13) | 5.42(47) | 3.853(47) | 3.853(47) | 3(12) |
| $R_{inner}$ (cm) | 3(18) | 3(18) | PP | 1(17) | 1.5(17) | 1.5(17) | PP |
| $T_{BB}$ (°K) | 20,000 | 40,000 | 40,000 | 150,000 | 75,000 | 75,000 | power law |
| Geometry | closed | closed | closed | closed | closed | closed | open |
| Stopping criteria | H-front | H-front | H-front | H-front | 7.5E17 cm | 7.5E17 cm | 1E22 cm$^{-2}$ |
| $n_H$ (cm$^{-3}$) | 100 | 100 | 1(4) | 3000 | 500 | 15,810 | 10,000 |
| He | 0.10 | 0.10 | 0.095 | 0.10 | 0.10 | 0.10 | 0.10 |
| C | 2.2(-4) | 2.2(-4) | 3.0(-4) | 3.0(-4) | 2.0(-4) | 2.0(-4) | 3.0(-4) |
| N | 4.0(-5) | 4.0(-5) | 7.0(-5) | 1.0(-4) | 6.0(-5) | 6.0(-5) | 1.0(-4) |
| O | 3.3(-4) | 3.3(-4) | 4.0(-4) | 6.0(-4) | 3.0(-4) | 3.0(-4) | 8.0(-4) |
| Ne | 5.0(-5) | 5.0(-5) | 1.1(-4) | 1.5(-4) | 6.0(-5) | 6.0(-5) | 1.0(-4) |
| Mg | | | | 3.0(-5) | 1.0(-5) | 1.0(-5) | 3.0(-5) |
| Si | | | | 3.0(-5) | 1.0(-5) | 1.0(-5) | 3.0(-5) |
| S | 9.0(-6) | 9.0(-6) | 1.0(-5) | 1.5(-5) | 1.0(-5) | 1.0(-5) | 1.5(-5) |
| Ar | | | | 3.0(-6) | | | |

**Ferland**: (2) Work on Cloudy began at the Institute of Astronomy, Cambridge, in 1978. Version 87 was used to generate the results presented here. A recent publication is Ferland, Fabian, & Johnston (1994). The code is fully described in the document *Hazy, a Brief Introduction to Cloudy*. Instructions for obtaining both the documentation and the code via anonymous ftp may be obtained from the author. For the models presented here outward-only continuum transfer was used, and the guesses of the dielectronic recombination rate coefficients described by Ali et al. (1991) were turned off.

**Harrington**: (3) These codes are discussed in Harrington (1968), Harrington et al. (1982), and Clegg et al. (1987). Most versions are for PNe; the radiation extends only to 200 eV. A version developed by S. Kraemer (1986) for AGNs goes to 3 keV. Most work has spherical geometry, though plane-parallel versions exist (Bregman & Harrington 1986). The spherical code used here (except for Table 4) starts with the outward-only approximation, followed by iterations in which the diffuse field is computed by Feautrier techniques from the opacity and source functions of the previous model. Convergence may take ~20 iterations for cases like the Meudon PN model (i.e., high optical depth and a hot star). Codes are not available as they are not documented and exist in many versions. A version ("nebmod") is in the UK Starlink collection, but documentation is minimal.

**Kallman**: (4) The XSTAR code dates from work at the University of Colorado by R. McCray and students: J. Buff, S. Hatchett, C. Wright, and R. Ross. The code was originally developed primarily for the prediction of X-Ray line emission from gas in X-Ray binaries, and later modified to treat optical and UV line emission from broad line clouds in quasars. A comprehensive description and collection of results were published in Kallman and McCray (1982). A more up-to-date description is included in the user's manual which, together with the source code, is available by anonymous ftp from legacy.gsfc.nasa.gov.

Table 2 Cool HII Region

|  |  | Mean | 1 | 2 | 3 | 5 | 6 | 8 | 9 | 10 | 11 |
|---|---|---|---|---|---|---|---|---|---|---|---|
| L(Hβ) | E36 | 4.93 | 4.99 | 4.98 | 4.93 | 4.85 | 4.83 | 4.93 | 4.94 | 5.01 | 4.91 |
| [NII] | 6584+ | 0.85 | 0.82 | 0.91 | 0.82 | 0.97 | 0.82 | 0.84 | 0.84 | 0.83 | 0.83 |
| [OII] | 3727+ | 1.18 | 1.11 | 1.16 | 1.22 | 1.32 | 1.14 | 1.21 | 1.24 | 1.14 | 1.11 |
| [NeII] | 12.8 μ | 0.31 | 0.36 | 0.35 | 0.29 | 0.29 | 0.29 | 0.29 | 0.35 | 0.29 | 0.29 |
| [SII] | 6720+ | 0.57 | 0.69 | 0.64 | 0.55 | 0.61 | 0.52 | 0.52 | 0.60 | 0.45 | 0.58 |
| [SIII] | 18.7 μ | 0.32 | 0.26 | 0.27 | 0.36 | 0.17 | 0.37 | 0.37 | 0.33 | 0.40 | 0.30 |
| [SIII] | 34 μ | 0.52 | 0.43 | 0.47 | 0.60 | 0.27 | 0.61 | 0.62 | 0.54 | 0.67 | 0.51 |
| [SIII] | 9532+ | 0.55 | 0.40 | 0.48 | 0.55 | 0.64 | 0.60 | 0.56 | 0.49 | 0.62 | 0.58 |
| L(total) | E36 | 21.2 | 20.3 | 21.3 | 21.7 | 20.7 | 21.0 | 21.8 | 21.7 | 22.1 | 20.6 |
| T(in) |  | 6793 | 6860 | 6952 | 6749 | 6980 | 6870 | 6747 | 6230 | 6912 | 6838 |
| T(H+) |  | 6744 | 6690 | 6740 | 6742 | 6950 | 6660 | 6742 | 6770 | 6720 | 6681 |
| <He+>/<H+> |  | 0.054 |  | 0.041 | 0.044 | 0.068 | 0.048 | 0.034 | 0.055 | 0.090 |  |
| R(out) | E18 | 8.96 | 9.00 | 8.93 | 8.94 | 9.00 | 8.93 | 9.00 | 8.87 | 9.02 | 8.97 |

**Netzer**: (5) ION was born in 1976 and reached some maturity in 1978. The earlier versions of the code are described in Netzer and Ferland (1984), Rees, Netzer & Ferland (1989) and Netzer (1993). ION is designed to cover the full range of densities from extremely low up to $10^{13}$ cm$^{-3}$. The HI, HeI and HeII solutions include a large number of levels and a full transfer treatment. Line transfer is calculated with the local escape probability and continuum transfer with either modified on-the-spot or outward only approximations. The present calculations assume outward-only and low temperature dielectronic recombination to magnesium and sulfur were not included. ION is available upon request from the author.

**Péquignot**: (6) NEBU is a descendant of a photoionization code whose main features were established by 1978 in a collaboration with G. Stasinska and S. Viegas at Meudon and São Paulo. Continuum and lines are treated outward only along 20 directions in spherical symmetry. NEBU is a combined photoionization and planar-shock code taking into account time-dependent ionization. The shock version was used by Stasinska and Péquignot at the Meudon meeting. Non-collisional excitation of forbidden lines was implemented in 1989 (Petitjean, Boisson & Péquignot, 1990; this includes a short description of the code). The codes AANGABA (Viegas) and PHOTO (Stasinska) diverged from NEBU in 1985 and 1988 respectively.

Table 3 Meudon HII Region

| | | Meu | Lex | 1 | 2 | 3 | 4 | 5 | 6 | 8 | 9 | 10 | 11 |
|---|---|---|---|---|---|---|---|---|---|---|---|---|---|
| L(Hβ) | E37 | 2.06 | 2.03 | 1.96 | 2.06 | 2.04 | 1.86 | 2.02 | 2.02 | 2.05 | 2.10 | 2.11 | 2.09 |
| HeI | 5876 | 0.116 | 0.116 | 0.125 | 0.109 | 0.119 | 0.110 | 0.101 | 0.116 | | 0.125 | 0.115 | 0.120 |
| CII | 2326+ | 0.17 | 0.16 | 0.07 | 0.19 | 0.17 | 0.16 | 0.16 | 0.14 | 0.18 | 0.28 | 0.12 | 0.14 |
| CIII] | 1909+ | 0.051 | 0.06 | 0.050 | 0.059 | 0.059 | 0.027 | 0.078 | 0.065 | 0.076 | 0.082 | 0.077 | 0.071 |
| [NII] | 122 μ | | 0.031 | 0.032 | 0.033 | | | | 0.036 | 0.031 | 0.030 | 0.037 | 0.034 |
| [NII] | 6584+ | 0.73 | 0.79 | 0.61 | 0.88 | 0.74 | 0.94 | 0.87 | 0.78 | 0.73 | 0.78 | 0.81 | 0.75 |
| [NIII] | 57 μ | 0.30 | 0.27 | 0.16 | 0.27 | 0.29 | | 0.26 | 0.30 | 0.30 | 0.17 | 0.27 | 0.39 |
| [OII] | 3727+ | 2.01 | 2.16 | 1.50 | 2.19 | 2.14 | 2.56 | 2.3 | 2.11 | 2.26 | 2.41 | 2.20 | 1.95 |
| [OIII] | 51.8 μ | 1.10 | 1.07 | 1.10 | 1.04 | 1.11 | 1.04 | 0.99 | 1.08 | 1.08 | 1.23 | 1.04 | 0.97 |
| [OIII] | 88.4 μ | 1.20 | 1.23 | 1.30 | 1.07 | 1.28 | | 1.16 | 1.25 | 1.26 | 1.42 | 1.20 | 1.14 |
| [OIII] | 5007+ | 2.03 | 2.06 | 2.30 | 1.93 | 1.96 | 1.47 | 2.29 | 2.17 | 2.10 | 2.23 | 2.22 | 1.89 |
| [NeII] | 12.8 μ | 0.21 | 0.22 | 0.26 | 0.23 | 0.19 | 0.23 | 0.22 | 0.20 | 0.20 | 0.22 | 0.22 | 0.20 |
| [NeIII] | 15.5 μ | 0.44 | 0.38 | 0.37 | 0.43 | 0.43 | 0.47 | 0.37 | 0.42 | 0.42 | 0.22 | 0.34 | 0.38 |
| [NeIII] | 3869+ | 0.096 | 0.086 | 0.085 | 0.103 | 0.086 | 0.071 | 0.100 | 0.079 | 0.087 | 0.081 | 0.087 | 0.078 |
| [SII] | 6720+ | 0.14 | 0.20 | 0.24 | 0.23 | 0.16 | 0.25 | 0.22 | 0.17 | 0.13 | 0.21 | 0.15 | 0.21 |
| [SIII] | 18.7 μ | 0.55 | 0.55 | 0.56 | 0.48 | 0.56 | 0.53 | 0.5 | 0.55 | 0.58 | 0.58 | 0.58 | 0.55 |
| [SIII] | 34 μ | 0.93 | 0.89 | 0.91 | 0.82 | 0.89 | | 0.81 | 0.88 | 0.94 | 0.92 | 0.92 | 0.91 |
| [SIII] | 9532+ | 1.25 | 1.29 | 1.16 | 1.27 | 1.23 | 1.15 | 1.48 | 1.27 | 1.30 | 1.31 | 1.32 | 1.46 |
| [SIV] | 10.5 μ | 0.39 | 0.34 | 0.22 | 0.37 | 0.42 | 0.35 | 0.36 | 0.41 | 0.33 | 0.26 | 0.38 | 0.27 |
| L(total) | E37 | 24.1 | 24.2 | 21.7 | 24.1 | 24.1 | 17.4 | 24.8 | 24.3 | 24.6 | 26.4 | 25.5 | 24.1 |
| T(in) | | | 7378 | 7552 | 7630 | 7815 | 7741 | 8057 | 7670 | 7650 | 7399 | 6530 | 7582 | 7445 |
| T(H+) | | | 7992 | 8034 | 7880 | 8064 | 8047 | 7879 | 8000 | 8060 | 8087 | 8220 | 8191 | 7913 |
| <He+>/<H+> | | | 0.77 | | 0.71 | 0.77 | 0.69 | 0.76 | 0.75 | 0.83 | 0.86 | 0.79 | 0.77 |
| R(out) | E18 | 1.45 | 1.48 | 1.43 | 1.46 | 1.46 | 1.61 | 1.47 | 1.46 | 1.46 | 1.46 | 1.49 | 1.47 |

Table 4 Blister HII Region

| | | Mean | 1 | 2 | 3 | 4 | 5 | 6 | 8 | 9 | 10 | 11 |
|---|---|---|---|---|---|---|---|---|---|---|---|---|
| I(Hβ) | | | 4.62 | 4.60 | 4.59 | 4.81 | 3.89 | 4.69 | 4.67 | 4.70 | 4.85 | 4.58 | 4.78 |
| HeI | 5876 | 0.12 | 0.12 | 0.13 | 0.11 | 0.11 | 0.12 | 0.12 | | 0.12 | 0.11 | 0.12 |
| CII | 2326+ | 0.18 | 0.06 | 0.14 | 0.20 | 0.30 | 0.10 | 0.15 | 0.23 | 0.35 | 0.11 | 0.16 |
| CII | 1335+ | 0.09 | .002 | 0.17 | 0.14 | 0.02 | 0.13 | 0.16 | | 0.01 | 0.02 | 0.13 |
| CIII] | 1909+ | 0.17 | 0.13 | 0.22 | 0.17 | 0.08 | 0.18 | 0.15 | 0.20 | 0.25 | 0.14 | 0.23 |
| [NII] | 6584+ | 0.87 | 0.67 | 0.58 | 0.94 | 1.48 | 0.74 | 0.90 | 0.87 | 0.92 | 0.82 | 0.83 |
| [NIII] | 57 μ | .031 | .032 | .035 | .033 | | .033 | .032 | .034 | .014 | .032 | .033 |
| [OII] | 7330+ | 0.12 | 0.06 | 0.10 | 0.13 | 0.19 | 0.09 | 0.12 | 0.14 | 0.15 | 0.08 | 0.10 |
| [OII] | 3727+ | 0.88 | 0.53 | 0.73 | 0.98 | 1.39 | 0.69 | 0.86 | 1.04 | 1.04 | 0.73 | 0.86 |
| [OIII] | 51.8 μ | 0.29 | 0.29 | 0.31 | 0.29 | 0.26 | 0.28 | 0.28 | 0.28 | 0.32 | 0.28 | 0.27 |
| [OIII] | 5007+ | 4.13 | 4.50 | 4.74 | 3.90 | 3.28 | 4.40 | 3.90 | 3.96 | 4.51 | 4.16 | 3.98 |
| [NeII] | 12.8 μ | 0.36 | 0.45 | 0.32 | 0.33 | 0.44 | 0.35 | 0.33 | 0.35 | 0.36 | 0.37 | 0.35 |
| [NeIII] | 15.5 μ | 0.98 | 0.93 | 1.24 | 1.07 | 1.09 | 0.96 | 1.04 | 1.00 | 0.59 | 0.92 | 0.97 |
| [NeIII] | 3869+ | 0.33 | 0.33 | 0.48 | 0.32 | 0.31 | 0.35 | 0.26 | 0.29 | 0.35 | 0.31 | 0.31 |
| [SIII] | 18.7 μ | 0.35 | 0.37 | 0.31 | 0.34 | 0.37 | 0.31 | 0.33 | 0.35 | 0.37 | 0.34 | 0.39 |
| [SIII] | 9532+ | 1.53 | 1.52 | 1.41 | 1.46 | 1.62 | 1.51 | 1.42 | 1.53 | 1.61 | 1.42 | 1.82 |
| [SIV] | 10.5 μ | 0.46 | 0.26 | 0.54 | 0.52 | 0.42 | 0.51 | 0.53 | 0.43 | 0.36 | 0.50 | 0.51 |
| I(total) | | 50.3 | 47.1 | 52.6 | 52.4 | 44.2 | 50.4 | 49.4 | 50.3 | 54.9 | 47.4 | 52.9 |
| T(in) | | 7989 | 8300 | 8206 | 7582 | | 8200 | 8200 | 7366 | 7740 | 8122 | 8189 |
| T(H+) | | 8263 | 8170 | 8324 | 8351 | | 8310 | 8200 | 8328 | 8220 | 8217 | 8250 |
| <He+>/<H+> | | 0.85 | | 0.94 | 0.78 | | 0.93 | 0.79 | 0.84 | 0.86 | 0.85 | 0.84 |
| ΔR | E17 | 2.96 | 2.90 | 2.88 | 3.08 | | 2.93 | 2.98 | 3.09 | 3.10 | 2.67 | 3.03 |

**Raymond:** (7) The shock code described by Raymond (1979) grew out of the one developed by Cox (1972), who gives a clear description of the physical processes and the numerical methods. Updated atomic rates are described in Cox and Raymond (1985), and an extensive grid of models is given in Hartigan, Raymond and Hartmann (1987). The code assumes steady flow, which simplifies the hydrodynamics to a calculation of the radiative cooling rate and the resulting compression. It computes the time-dependent ionization balance including photoionization and the associated heating.

**Rubin**: (8) This code "NEBULA" has a detailed treatment for the Lyman continuum radiative transport. The gas structure (ionization and thermal equilibrium) and transfer of the diffuse ionizing radiation are solved iteratively. For the example H II region models here, uniform convergence is obtained in 15 - 20 iterations. A publication that uses a recent version of the 1-dimensional (spherical) code is by Simpson et al. (1995). A 2-dimensional, axisymmetric version, which treats the diffuse radiation in a less precise fashion, has been applied to model the Orion Nebula (Rubin et al. 1991; 1993).

**Shields**: (9) This code, called NEBULA, was first developed at Caltech in 1971 for work on Seyfert galaxies. It was later streamlined to include only physics relevant to H II regions and planetary nebulae. Some description is contained in Garnett and Shields (1987), and further details may be obtained from Shields. The models here were computed with OTS for the Lyman continua of H I and He II, outward-only for the Lyman continuum of He I, and outward-only for the helium Lyman lines.

Table 5 Meudon Planetary Nebula

|  |  | Meu | Lex | 1 | 2 | 3 | 4 | 5 | 6 | 10 | 11 |
|---|---|---|---|---|---|---|---|---|---|---|---|
| L(Hβ) | E35 | 2.60 | 2.53 | 2.06 | 2.63 | 2.68 | 2.35 | 2.73 | 2.68 | 2.30 | 2.80 |
| He I | 5876 | 0.11 | 0.09 | 0.09 | 0.11 | 0.10 | 0.05 | 0.10 | 0.11 | 0.09 | 0.11 |
| He II | 4686 | 0.33 | 0.41 | 0.40 | 0.32 | 0.33 | 0.81 | 0.35 | 0.32 | 0.43 | 0.34 |
| C II] | 2326+ | 0.38 | 0.27 | 0.20 | 0.33 | 0.43 | 0.12 | 0.27 | 0.30 | 0.22 | 0.32 |
| C III] | 1909+ | 1.70 | 2.14 | 2.40 | 1.82 | 1.66 | 2.92 | 1.72 | 1.87 | 3.14 | 1.63 |
| C IV | 1549+ | 1.64 | 2.51 | 2.60 | 2.44 | 2.05 | 1.44 | 2.66 | 2.18 | 4.74 | 1.94 |
| [N II] | 6584+ | 1.44 | 1.49 | 1.43 | 1.59 | 1.45 | 1.69 | 1.47 | 1.44 | 1.47 | 1.38 |
| N III] | 1749+ | 0.11 | 0.12 | 0.16 | 0.13 | 0.13 | 0.01 | 0.11 | 0.13 | 0.16 | 0.16 |
| [NIII] | 57 μ |  | 0.13 | 0.11 | 0.12 | 0.13 |  | 0.13 | 0.13 | 0.13 | 0.14 |
| N IV] | 1487+ | 0.12 | 0.20 | 0.22 | 0.20 | 0.15 | 0.20 | 0.21 | 0.19 | 0.26 | 0.17 |
| N V | 1240+ | 0.09 | 0.21 | 0.17 | 0.18 | 0.12 | 0.34 | 0.23 | 0.15 | 0.38 | 0.15 |
| [O I] | 6300+ | 0.15 | 0.14 | 0.17 | 0.15 | 0.12 | 0.15 | 0.14 | 0.14 | 0.16 | 0.13 |
| [O II] | 3727+ | 2.23 | 2.28 | 2.35 | 2.23 | 2.27 |  | 2.31 | 2.18 | 2.50 | 2.14 |
| [O III] | 5007+ | 20.9 | 20.7 | 21.8 | 21.1 | 21.4 | 19.8 | 19.4 | 21.1 | 20.2 | 20.9 |
| [O III] | 4363 | 0.16 | 0.16 | 0.18 | 0.16 | 0.16 | 0.19 | 0.14 | 0.16 | 0.16 | 0.15 |
| [O III] | 52 μ | 1.43 | 1.34 | 1.39 | 1.42 | 1.44 | 0.96 | 1.40 | 1.46 | 1.26 | 1.41 |
| [O IV] | 26 μ | 3.62 | 3.92 | 3.90 | 3.52 | 3.98 | 4.48 | 3.32 | 3.86 | 5.01 | 3.33 |
| O IV] | 1403+ | 0.13 | 0.27 | 0.36 | 0.20 | 0.23 | 0.18 | 0.26 | 0.33 | 0.41 | 0.15 |
| O V] | 1218+ | 0.09 | 0.24 |  | 0.20 | 0.11 | 0.35 | 0.29 | 0.19 | 0.33 |  |
| [Ne III] | 15.5 μ | 2.51 | 2.49 | 2.67 | 2.75 | 2.76 | 0.72 | 2.80 | 2.81 | 2.71 | 2.74 |
| [Ne III] | 3869+ | 2.59 | 2.63 | 3.20 | 3.33 | 2.27 | 0.88 | 2.74 | 2.44 | 3.35 | 2.86 |
| Ne IV] | 2423+ | 0.56 | 0.95 | 1.05 | 0.72 | 0.74 | 1.64 | 0.91 | 0.74 | 1.19 | 0.63 |
| [Ne V] | 3426+ | 0.73 | 0.90 | 0.79 | 0.74 | 0.60 | 2.29 | 0.73 | 0.61 | 0.81 | 0.63 |
| [Ne V] | 24.2 μ | 1.67 | 0.88 | 1.20 | 0.94 | 0.76 | 1.16 | 0.81 | 0.99 | 0.25 | 0.95 |
| Mg II | 2798+ | 1.48 | 1.56 | 2.50 | 2.33 | 1.60 | 0.63 | 1.22 | 1.17 | 1.15 | 1.92 |
| [Mg IV] | 4.5 μ | 0.09 | 0.12 | 0.11 | 0.12 | 0.13 |  |  | 0.12 | 0.14 | 0.12 |
| [Si II] | 34.8 μ | 0.13 | 0.18 | 0.14 | 0.16 | 0.26 |  | 0.19 | 0.17 | 0.15 | 0.16 |
| Si II] | 2335+ | 0.11 | 0.23 | 0.23 | 0.15 |  | 0.53 | 0.16 | 0.16 |  | 0.15 |
| Si III] | 1892+ | 0.20 | 0.68 | 0.79 | 0.39 | 0.32 | 1.95 | 0.46 | 0.45 |  | 0.40 |
| Si IV | 1397+ | 0.15 | 0.15 | 0.10 | 0.20 | 0.15 | 0.03 | 0.21 | 0.17 |  | 0.16 |
| [S II] | 6720+ | 0.39 | 0.33 | 0.24 | 0.21 | 0.45 | 0.08 | 0.33 | 0.43 | 0.41 | 0.51 |
| [S III] | 18.7 μ | 0.49 | 0.53 | 0.60 | 0.48 | 0.49 |  | 0.46 | 0.49 | 0.60 | 0.55 |
| [S III] | 9532+ | 2.09 | 1.91 | 2.31 | 2.04 | 1.89 | 0.36 | 2.05 | 1.87 | 2.34 | 2.42 |
| [S IV] | 10.5 μ | 1.92 | 1.84 | 1.58 | 1.92 | 2.21 | 0.93 | 1.81 | 1.98 | 2.36 | 1.94 |
| L(total) | E35 | 129 | 132 | 114 | 139 | 136 | 105 | 135 | 136 | 130 | 142 |
| T(in) | E4 |  | 1.80 |  | 1.83 | 1.78 | 1.63 | 1.84 | 1.78 | 1.95 | 1.81 |
| T(H+) | E4 |  | 1.26 |  | 1.22 | 1.21 | 1.32 | 1.35 | 1.21 | 1.29 | 1.20 |
| <He+>/<H+> |  |  | 0.69 |  | 0.74 | 0.74 |  | 0.71 | 0.71 | 0.60 | 0.68 |
| R(out) | E17 |  | 4.02 |  | 4.04 | 4.04 |  | 4.07 | 4.07 | 3.83 | 4.08 |

**Sutherland**: (10) The MAPPINGS code was started in 1976 by Dopita to model plane parallel steady shocks. In 1990 a rewrite was performed to extend MAPPINGS into the high temperature regime (Sutherland and Dopita 1993). MAPPINGS II, does not supersede the Binette MAPPINGS code but is rather an extension in a different direction. MAPPINGS II v1.0.5x was used for the results presented here. The diffuse field is calculated and integrated in the outward only approximation. Dielectronic recombination rates for sulfur and argon based on the Opacity Project cross-sections have been used for Ar I-V and S I-V. Inquiries about the availability of MAPPINGS II should be made to the author.

**Viegas**: (11) AANGABA (which means "image" in the Brazilian native language) is a descendant of a photoionization code whose main features were established by 1978 in a collaboration with G. Stasinska and D. Péquignot at Meudon and São Paulo. The diffuse radiation field is calculated using the outward-only approximation along 20 directions in spherical symmetry. A brief description of the code is given in Gruenwald & Viegas (1992). The codes NEBU (Péquignot) and PHOTO (Stasinska) diverged from AANGABA in 1985.

**Viegas and Contini**: (12) The first version of SUMA, which accounts for the coupled effects of photoionization and shocks, appeared in 1982. The code has been revised in 1992, and the photoionization calculation is in full agreement with AANGABA. The last publication refers to results obtained after updating the code and including the effect of dust (Viegas & Contini 1994).

Table 6 High Ionization PN

|   |   | Mean | 1 | 2 | 3 | 4 | 5 | 6 | 10 | 11 |
|---|---|------|---|---|---|---|---|---|----|----|
| L(Hβ) | E34 | 5.85 | 5.67 | 6.05 | 5.96 | 6.02 | 5.65 | 5.74 | 5.72 | 6.02 |
| He I | 5876 | 0.12 | 0.12 | 0.13 | 0.13 | 0.10 | 0.12 | 0.13 | 0.13 | 0.13 |
| He II | 4686 | 0.081 | 0.096 | 0.080 | 0.087 | 0.039 | 0.085 | 0.092 | 0.090 | 0.083 |
| C III] | 1909+ | 0.83 | 0.90 | 0.60 | 0.60 | 0.89 | 0.99 | 0.89 | 1.03 | 0.74 |
| C IV | 1549+ | 0.34 | 0.24 | 0.35 | 0.29 | 0.45 | 0.40 | 0.37 | 0.32 | 0.29 |
| [N II] | 6584+ | 0.12 | 0.12 | 0.11 | 0.11 | 0.14 | 0.15 | 0.12 | 0.12 | 0.12 |
| [NIII] | 57 μ | 0.39 | 0.27 | 0.37 |  |  | 0.40 | 0.41 | 0.40 | 0.48 |
| [O II] | 3727+ | 0.29 | 0.32 | 0.22 | 0.24 | 0.35 | 0.35 | 0.26 | 0.32 | 0.27 |
| [O III] | 5007+ | 11.5 | 12.1 | 10.0 | 10.1 | 12.7 | 12.2 | 11.7 | 11.9 | 11.2 |
| [O III] | 52 μ | 2.02 | 2.03 | 1.88 | 1.96 | 2.39 | 1.95 | 2.02 | 2.02 | 1.94 |
| [O IV] | 26 μ | 0.79 | 0.76 | 0.68 | 0.80 | 1.09 | 0.71 | 0.86 | 0.77 | 0.67 |
| [Ne III] | 15.5 μ | 1.35 | 1.35 | 1.30 | 1.32 | 1.55 | 1.30 | 1.35 | 1.34 | 1.31 |
| [Ne III] | 3869+ | 1.03 | 1.15 | 1.02 | 0.92 | 1.02 | 1.13 | 0.89 | 1.11 | 1.00 |
| Mg II | 2798+ | 0.13 | 0.34 | 0.10 | 0.07 | 0.14 | 0.05 | 0.10 | 0.10 | 0.11 |
| Si III] | 1892+ | 0.15 | 0.10 | 0.09 | 0.10 | 0.37 | 0.15 | 0.13 |  | 0.11 |
| [S III] | 18.7 μ | 0.34 | 0.45 | 0.24 | 0.32 |  | 0.26 | 0.28 | 0.36 | 0.49 |
| [S III] | 9532+ | 1.13 | 1.40 | 0.81 | 0.92 |  | 1.02 | 0.85 | 1.10 | 1.77 |
| [S IV] | 10.5 μ | 2.05 | 1.66 | 2.19 | 2.21 | 2.10 | 2.11 | 2.35 | 2.31 | 1.46 |
| L(total) | E34 | 133 | 133 | 122 | 120 | 140 | 132 | 131 | 134 | 134 |
| T(in) | E4 | 1.48 | 1.45 | 1.48 | 1.42 | 1.83 | 1.40 | 1.45 | 1.36 | 1.44 |
| T(H+) | E4 | 1.05 | 1.07 | 1.01 | 1.01 | 1.03 | 1.14 | 1.05 | 1.06 | 1.03 |
| <He+>/<H+> |  | 0.92 |  | 0.92 | 0.92 |  | 0.92 | 0.91 | 0.92 | 0.92 |

Table 7 Low Ionization PN

|         |       | Mean  | 1     | 2     | 3     | 5     | 6     | 10    | 11    |
|---------|-------|-------|-------|-------|-------|-------|-------|-------|-------|
| L(Hβ)   | E34   | 5.41  | 5.20  | 5.56  | 5.52  | 5.35  | 5.41  | 5.38  | 5.44  |
| He I    | 5876  | 0.13  | 0.12  | 0.14  | 0.12  | 0.14  | 0.12  | 0.12  | 0.15  |
| He II   | 4686  | .088  | .095  | .082  | .088  | .088  | .088  | .091  | .086  |
| C III]  | 1909+ | 1.17  | 1.53  | 0.78  | 0.81  | 1.25  | 1.41  | 1.43  | 1.00  |
| C IV    | 1549+ | 1.38  | 1.20  | 1.34  | 1.31  | 1.54  | 1.43  | 1.51  | 1.32  |
| [O III] | 5007+ | 14.3  | 16.0  | 12.7  | 13.1  | 15.0  | 14.5  | 14.6  | 14.3  |
| [O III] | 52 μ  | 0.26  | 0.28  | 0.26  | 0.26  | 0.26  | 0.27  |       | 0.26  |
| [O IV]  | 26 μ  | 0.22  | 0.24  | 0.21  | 0.23  | 0.21  | 0.21  | 0.22  | 0.22  |
| [Ne III]| 15.5 μ| 1.11  | 1.14  | 1.09  | 1.10  | 1.10  | 1.12  | 1.12  | 1.10  |
| [Ne III]| 3869+ | 1.44  | 1.66  | 1.39  | 1.27  | 1.48  | 1.44  | 1.45  | 1.38  |
| Ne IV]  | 2423+ | 0.10  | 0.10  | 0.11  | 0.11  | 0.10  | 0.08  | 0.08  | 0.10  |
| [S III] | 9532+ | 0.52  | 0.65  | 0.23  | 0.37  | 0.26  | 0.42  | 0.76  | 0.99  |
| [S IV]  | 10.5 μ| 1.43  | 1.20  | 1.32  | 1.57  | 1.32  | 1.84  | 1.38  | 1.37  |
| L(total)| E34   | 120   | 126   | 109   | 112   | 122   | 124   | 122   | 121   |
| T(in)   | E4    | 1.78  |       | 1.83  | 1.79  | 1.76  | 1.73  | 1.76  | 1.81  |
| T(H+)   | E4    | 1.16  |       | 1.11  | 1.11  | 1.28  | 1.14  | 1.20  | 1.14  |
| <He+>/<H+> |    | 0.91  |       | 0.91  | 0.91  | 0.91  | 0.91  | 0.90  | 0.90  |

# 3. The Model Nebulae

These benchmarks are designed to be simple yet still test the numerical approximations used in the simulations. Blackbodies are used instead of stellar atmospheres; the compositions have low gas phase abundances for refractory elements but do not include the grains themselves; and all have constant density. For shocks, the gradual liberation of refractory elements in the course of grain destruction (see Jones et al. 1994) and departures from steady flow (especially for $v_s$ 8 150 km/s; see Innes 1992) are ignored, as are the photoionization precursors which contribute significantly to the optical emission lines if one observes an entire SNR rather than a single bright filament (see Shull 1983).

## 3.1 Cloud Parameters and Tabulated Quantities

Table 1 lists the parameters for the benchmarks presented here. The first row gives the table number for the models, and following rows list parameters. Standard notation is used (Osterbrock 1989).

The luminosity or intensity of the continuum is specified by a) for the spherical geometry, Q(H), the total number of hydrogen-ionizing photons emitted per second into 4π sr; or b) for the plane-parallel case, the flux of hydrogen ionizing photons at the illuminated face of the cloud . The inner radius is given for spherical geometries, or the notation "PP" if the geometry is plane parallel. The black body temperature follows, except for model 8 which uses the power-law described below. Most models assumed a "closed" geometry, one in which radiation escaping from the illuminated face of the cloud does not escape the system. An open geometry is one in which such radiation freely escapes.

The stopping criteria follow. There are either the hydrogen ionization front ("H-front"), a certain thickness ΔR, or a column density $N_H$. The hydrogen density $n_H$ is listed, followed by abundances by number relative to hydrogen.

## 3.2 Results

The tables present several types of results for the models. The most important (observationally) are the intensities of lines - these are presented relative to Hβ. For multiplets the total intensities of all lines are listed; and this is indicated by the "+" after the wavelength of the multiplet. The total intensity or luminosity ("I(total)" or "L(total)") is the intensity or luminosity of Hβ (the first line) multiplied by the sum of the listed intensities. This is a measure of the energy conservation of the codes. The electron temperature at the illuminated face of

Table 8 NLR Cloud

|         |        | Mean | 1    | 2    | 4    | 5    | 6    | 11   |
|---------|--------|------|------|------|------|------|------|------|
| I(Hβ)   | E0     | 1.31 | 1.33 | 1.31 | 1.06 | 1.37 | 1.43 | 1.34 |
| Lyα     | 1216   | 34.2 | 38.3 | 32.1 | 37.0 | 32.4 | 31.5 | 34.2 |
| HeI     | 5876   | 0.12 | 0.11 | 0.13 | 0.14 | 0.12 | 0.13 | 0.13 |
| HeII    | 4686   | 0.24 | 0.25 | 0.25 |      | 0.25 | 0.23 | 0.24 |
| HeII    | 1640   | 1.60 | 1.60 | 1.74 | 1.49 | 1.53 | 1.56 | 1.67 |
| CIII]   | 1909+  | 2.82 | 2.90 | 2.99 | 2.45 | 2.87 | 2.83 | 2.90 |
| CIV     | 1549+  | 3.18 | 2.70 | 3.85 | 2.28 | 3.69 | 3.17 | 3.36 |
| [NII]   | 6584+  | 2.33 | 1.40 | 3.20 | 1.21 | 3.10 | 2.67 | 2.40 |
| NIII]   | 1749+  | 0.19 | 0.24 | 0.24 | 0.01 | 0.22 | 0.22 | 0.22 |
| NIV]    | 1487+  | 0.20 | 0.20 | 0.23 | 0.12 | 0.22 | 0.21 | 0.21 |
| [OI]    | 6300+  | 1.61 | 2.20 | 1.61 | 1.41 | 1.67 | 1.31 | 1.46 |
| [OI]    | 63 μ   | 1.12 | 0.25 | 1.13 |      |      | 1.44 | 1.64 |
| [OII]   | 3727+  | 1.72 | 1.60 | 1.44 | 3.18 | 1.58 | 1.30 | 1.20 |
| OIII]   | 1663+  | 0.56 | 0.35 | 0.63 |      | 0.61 | 0.57 | 0.63 |
| [OIII]  | 5007+  | 33.1 | 31.4 | 34.5 | 31.1 | 33.0 | 32.8 | 36.0 |
| [OIII]  | 4363   | 0.32 | 0.30 | 0.34 |      | 0.31 | 0.30 | 0.33 |
| OIV     | 1403+  | 0.36 | 0.49 | 0.30 |      | 0.36 | 0.42 | 0.25 |
| [NeIII] | 15.5 μ | 1.89 | 1.50 | 2.01 |      | 1.94 | 2.05 | 1.95 |
| [NeIII] | 3869+  | 1.91 | 1.90 | 2.51 | 0.84 | 2.16 | 1.72 | 2.34 |
| [Ne IV] | 2423+  | 0.44 | 0.52 | 0.42 |      | 0.47 | 0.41 | 0.38 |
| [NeV]   | 3426+  | 0.52 | 0.59 | 0.55 |      | 0.53 | 0.44 | 0.50 |
| MgII    | 2798+  | 1.78 | 3.50 | 1.72 | 1.48 | 1.23 | 1.12 | 1.61 |
| [SiII]  | 34.8 μ | 0.90 | 1.00 | 0.96 |      | 1.07 | 0.96 | 0.52 |
| [SII]   | 6720+  | 1.33 | 2.40 | 1.01 | 1.58 | 0.93 | 0.99 | 1.10 |
| [SIII]  | 9532+  | 1.88 | 1.60 | 2.15 | 1.73 | 2.06 | 1.67 | 2.08 |
| [SIII]  | 18.7 μ | 0.49 | 0.36 | 0.61 |      | 0.57 | 0.52 | 0.37 |
| [SIV]   | 10.5 μ | 1.05 | 0.86 | 1.24 | 1.23 | 0.82 | 0.94 | 1.22 |
| I(total)| E0     | 125  | 131  | 128  | 92   | 128  | 131  | 133  |
| T(in)   | E4     | 1.70 | 1.71 | 1.70 |      | 1.72 | 1.68 | 1.68 |
| T(H+)   | E4     | 1.17 |      | 1.24 | 1.12 | 1.06 | 1.20 | 1.23 |

the cloud is next, followed by the mean temperature T(H+), and the mean He⁺/H⁺:

$$T(H^+) = \frac{\int n_e n_p T_e \, dV}{\int n_e n_p \, dV}, \qquad \frac{\langle He^+ \rangle}{\langle H^+ \rangle} = \frac{n(H)}{n(He)} \frac{\int n_e n(He^+) \, dV}{\int n_e n_p \, dV}.$$

Table 2. This is an HII region ionized by a very cool star. This is the simplest model since helium is predominantly neutral.

Table 3. This was one of the original Meudon tests. It is a spherical HII region with a star warm enough to nearly fully fill the H⁺ Strömgren sphere with He⁺. Atomic helium radiative transfer is important, but the complexities introduced by fully ionized helium are not yet present. In this and the following tables the entries "Meu" and "Lex" indicate means for the Meudon and Lexington results.

Table 4. This model is motivated by conditions in the innermost regions of the Orion nebula. It is plane parallel, the simplest representation of the flow off the face of a molecular cloud.

Table 5. This is the original Meudon PN. These calculations are very sensitive to the Bowen OIII HeII Lyα fluorescence problem since helium is ionized in much of the nebula (see Netzer and Ferland 1984).

Tables 6 and 7. These PNe have the same intermediate temperature central star and chemical composition. The conditions are simpler than the Meudon planetary because the transfer of helium diffuse fields does not dominate the nebula. Table 6 has a low density and resulting high ionization. The high density PN (Table 7) has low ionization. Both have outer radii near the H-ionization front. The models are density bounded and so have a simpler ionization structure than the Meudon PN.

Table 8. This AGN Narrow Line Region model is irradiated by a nonthermal continuum with $f_\nu \propto \nu^{-1.3}$ between 1.36 eV and 50 keV, and no radiation outside this range. This tests high-energy effects, but the results are not sensitive to line transfer since emission lines are well below their thermalization density and Lyα trapping is not important.

Table 9 Shocks

|        |       | 6     | 7     | 10    | 12   |
|--------|-------|-------|-------|-------|------|
| I(Hβ)  | E–6   | 6.17  | 4.24  | 4.74  | 2.60 |
| Hα     | 6563  | 3.34  | 3.16  | 3.18  | 3.80 |
| H 2-ν  |       |       | 12.8  | 19.1  | 16.2 | - |
| He I   | 5876  | 0.15  | 0.14  | 0.10  | 0.18 |
| HeII   | 1640  | 2.14  | 0.33  |       |      |
| C II]  | 2326+ | 2.54  | 3.50  | 2.47  | 1.97 |
| [C II] | 158μ  | 0.76  | 1.23  | 0.73  | 1.06 |
| C III] | 1909+ | 6.87  | 8.83  | 6.03  | 8.10 |
| C IV   | 1549+ | 8.31  | 7.83  | 19.3  | 18.1 |
| [N II] | 6584+ | 1.48  | 1.91  | 1.96  | 1.43 |
| N III] | 1749+ | 0.75  | 0.90  | 0.75  | 0.87 |
| N IV]  | 1487+ | 0.26  | 0.43  | 0.23  | 0.54 |
| N V    | 1240+ | 0.02  | 0.02  | 0.03  | 0.12 |
| [O I]  | 6300+ | 0.07  | 0.16  | 0.17  | 0.03 |
| [OI]   | 63.2μ | 0.40  | 0.24  |       |      |
| [O II] | 3727+ | 6.59  | 8.69  | 10.4  | 10.4 |
| [O III]| 5007+ | 5.88  | 6.30  | 3.98  | 5.54 |
| O III] | 1663+ | 2.06  | 2.42  | 1.62  | 3.11 |
| O IV]  | 1403+ | 0.54  | 1.11  | 0.98  | 0.98 |
| [NeII] | 12.8μ | 0.40  | 0.33  | 0.41  |      |
| [NeIII]| 3968+ | 0.63  | 0.68  | 0.68  | 0.56 |
| NeIV]  | 2423+ | 0.06  | 0.27  | 0.07  | 0.24 |
| Mg II  | 2798+ | 0.25  | 0.52  | 0.54  | 0.06 |
| [Si II]| 34.8μ | 4.36  | 3.95  | 5.25  | 4.28 |
| SiIV]  | 1397+ | 2.31  | 2.43  |       |      |
| [S II] | 6720+ | 1.22  | 1.97  | 1.51  | 1.00 |
| [SIII] | 9532+ | 0.61  | 0.44  | 0.77  |      |
| [ArIII]| 7136+ | 0.22  | 0.22  | 0.14  |      |
| [CaII] | 7292+ | 0.14  | 0.14  | 0.20  | 0.04 |
| [Fe II]| sum   | 11.1  | 2.83  | 2.72  | 1.05 |
| ΔR     | E16   |       | 4.44  | 4.15  | 3.41 |

Table **9**. This table compares 100 km s$^{-1}$ steady-flow shock models. The pre-shock density and magnetic field are 1.0 cm$^{-3}$ and 1 µG, and the pre-shock gas is assumed to be fully ionized. Calculations stopped at 10$^3$ °K. The Allen (1973) abundances are assumed for the 13 to 15 most abundant elements through iron. Line intensities are given relative to Hβ, and the Hβ flux is given in erg cm$^{-2}$ s$^{-1}$, emitted into 4π sr. "[FeII] sum" is the sum of lines of the lowest 16 levels of the atom. The line "ΔR" is the depth in cm at which the gas cooled to 10$^3$ °K.

## 4. Outstanding Issues

Discussion within the meeting highlighted several outstanding issues, which are summarized here.

### 4.1 Radiation Transport

Two limiting cases of continuum transport exist: the "outward only" and "on-the-spot" approximations. Most codes use one or the other. Two (Harrington and Rubin) solve the problem correctly. Comparisons suggested that outward only gave better agreement with the exact solutions, especially with the temperature of the illuminated face of the cloud. Outward only was used for most calculations presented here.

### 4.2 The BLR

Models of the Broad Line Region of Active Galactic Nuclei were considered during the workshop, but no consensus was reached in the time available. These are challenging since the gas is dense enough for line transport and thermalization to be important, thus adding an extra layer of complexity on top of those present in conventional nebulae.

### 4.3 The Atomic Database

The Opacity Project (Seaton 1987) and other large-scale efforts (Verner and Yakovlev 1995) have resulted in a large homogeneous photoionization data base. The corresponding recombination (especially dielectronic) coefficients have been computed for second row elements, but do not yet exist for third row and iron group elements.

Photoionization models considered here do not include iron, which can dominate the thermal equilibrium of lower ionization clouds (Wills, Netzer, & Wills 1985). Work is now underway within the atomic physics community to generate the basic data needed to treat FeII emission properly.

## 4.4 Reliability — Photoionization Calculations

This can be judged by examining the scatter among the predictions. Figure 1 shows the relative standard deviation (the standard deviation divided by the mean) for all results presented for the Meudon HII region case (Table 3, the open circles). This is among the simplest cases since $He^{++}$ is not present. The dispersion is plotted against line excitation energy in inverse microns.

Our goal is for the simulations to agree within roughly 10%. This is the realistic limit to the *absolute* accuracy a calculation can hope for since modern calculations of theoretical atomic cross sections and rate coefficients are no more accurate than 10% - 15%. The lines with the largest standard deviation are largely those of third row elements, especially [SIII] and [SII]. Currently there is a large dispersion in photoionization cross sections for these many-electron systems, and no *total* recombination coefficients now exist (Ali et al. 1991).

The Meudon Planetary Nebula (Table 5) presents a more challenging case — a hot star, so that HeII transport problems are present. Results for all calculations are presented as the open circles in Figure 2. The dispersion is À30% with some lines larger than 50%, in whole disappointingly large.

The scatter in Figure 2 is most likely the result of different levels of treatment of the He line and continuum transport, a difficult problem (Osterbrock 1989, pp.

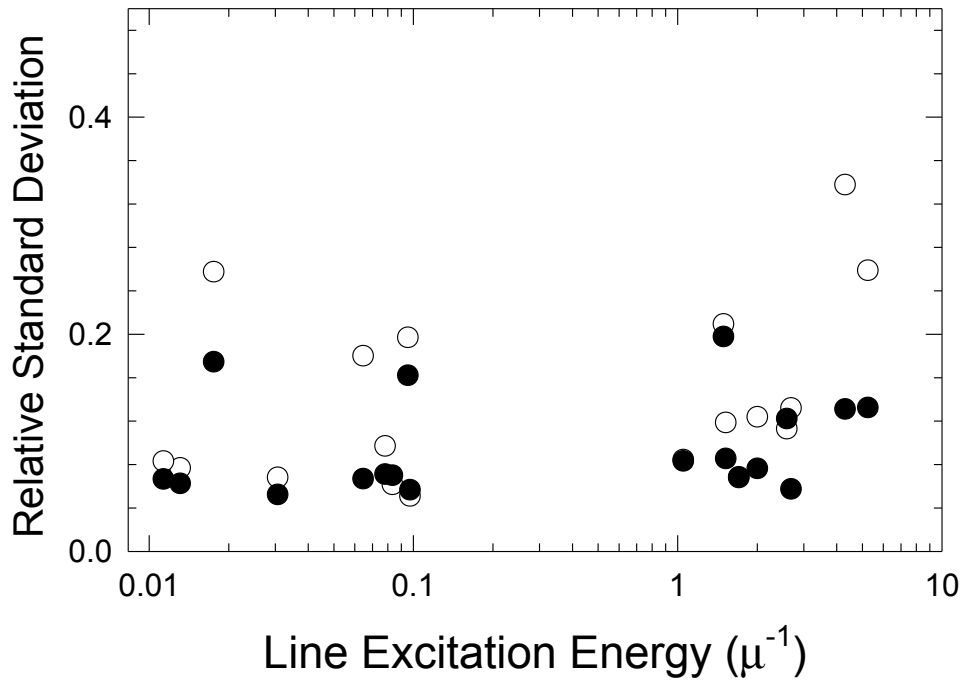

**Figure 1** Relative standard deviation (standard deviation divided by the mean) for the Meudon HII Region results (Table 3). The results for all calculations are shown as open circles, and the filled circles show those for the subset of the calculations described in the text.

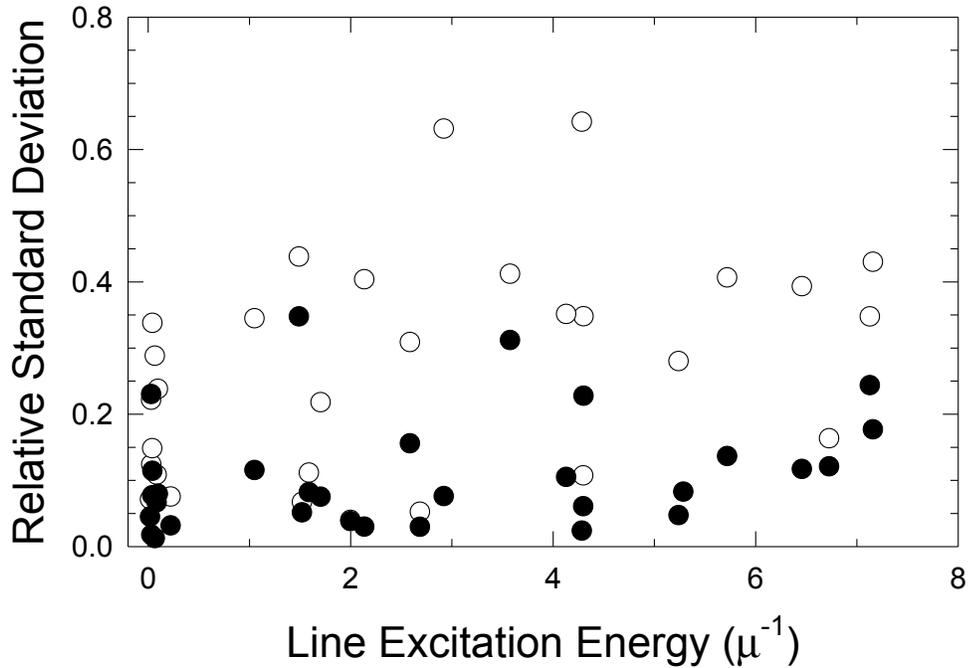

**Figure 2** Relative standard deviation for the Meudon PN results (Table 5). The results for all calculations are presented as open circles, and those for the calculations which fulfill the Hβ luminosity criterion (see text) by the filled circles.

23-31). One way to judge the accuracy of the overall transport problem is the luminosity of Hβ. For the Meudon PN the 100% conversion of ionizing photons into hydrogen recombinations (Table 1; Osterbrock 1989) would result in an Hβ luminosity of 2.58%10$^{35}$ erg s$^{-1}$. The actual Hβ luminosity will be larger, since the Bowen OIII process degrades HeII Lyα into several softer photons. We computed a second relative standard deviation for all lines, including only those calculations with predicted Hβ luminosities close to or larger than the 100% conversion value. These results, the predictions of codes 2, 3, 5, 6, and 11 are shown as filled circles in Figure 2. The mean Hβ intensity for these calculations is 2.70)0.06×10$^{36}$ erg s$^{-1}$. The scatter for this subset of the calculations is À10%. Clearly the computed spectrum is sensitive to the transport of the ionizing continuum.

We also recomputed a second relative standard deviation for the Meudon HII Region, including only the five calculations listed above, plus code 8. (L(Hβ) and L(total) for code 8 are well within 2 standard deviations of the five previous ones.) The second dispersion is shown on Figure 1 as the filled circles. Clearly, our stated goal of 10% accuracy can be realized.

## 4.5 Reliability — Shock Calculations

The most important predictions of the shock models are the intensities of the strong lines relative to Hβ, as these are used to infer shock velocities and elemental abundances. The scatter among the models is largest for lines from the very highest and very lowest ionization states. The former are most sensitive to ionization rates and conditions close to the shock, while the latter are most sensitive to details of the radiative transfer and thermal balance in the recombining gas. The discrepancies between the high ionization line predictions would only lead to À5 km/s difference between the shock velocities derived. The prediction of Hβ in shocks depends in a complex manner on collisional rates and diffuse radiation transfer, and so is not as simple to predict as in a photoionized nebula.

# 5. Conclusions

A major goal of this workshop was to establish a set of benchmark nebulae for reference in code development. These are presented in the tables throughout this chapter.

Intercomparisons among the calculations show that an overall spectroscopic accuracy approaching 10% can be realized, but this will require improvements in the atomic data base (especially recombination coefficients of third row elements) and continuum transport. The photoionization/recombination data base now constitutes a major uncertainty for calculations of third row elements. The situation is far better for second row elements, where this 10% accuracy is now realized.

**Acknowledgment:** The Lexington Meeting on Numerical Simulations of Ionized Nebulae was sponsored by the Graduate School, and the Center for Computational Sciences, of the University of Kentucky. We thank the Director of the CCS, Dr. John Connolly, for his hospitality.

# 6. References

Ali, B., et al., 1991, Publ. A.S.P. 103, 1182.
Allen, C.W., 1973, *Astrophysical Quantities* (London: Athlone Press).
Binette, L., Dopita, M., & Tuohy, I.R., 1985, ApJ, 297, 476.
Binette, L., Wang, J., Villar-Martin, M., Martin, P.G., & Magris C., G., 1993a, ApJ 414, 535.
Binette, L., Wang, J., Zuo, L., & Magris C., G., 1993b, AJ 105, 797.
Bregman, J, & Harrington, J., 1986, Ap.J.,309, 833.
Clegg, R.E.S., Harrington, J., Barlow, M., & Walsh, J.R., 1987: Ap.J., 314, 551
Cox, D.P., 1972 ApJ, 178, 143.
Cox, D.P., & Raymond, J.C. 1985, ApJ, 298, 651.
Draine, B.T., & McKee, C.F. 1993, Ann. Revs. Astr. Ap., 31, 373.

Ferland, G.J., Fabian, A.C., & Johnston, R., 1994, MNRAS 266, 399.
Garnett, D., & Shields, G.A., 1987, ApJ 317, 82.
Gruenwald, R.B., & Viegas, S., 1992. ApJS 78, 153.
Hartigan, P., Raymond, J., & Hartmann, L. 1987, ApJ, 316, 323.
Harrington, J.P., 1968, Ap.J.,152,943
Harrington, J.P., Seaton, M.J., Adams, S., & Lutz, J. 1982, MNRAS,199,517
Innes, D.E. 1992, A& A, 256, 660.
Jones, A.P., Tielens, A.G.G.M., Hollenbach, D.J., & McKee, C.F. 1994, Ap.J. 433, 797.
Kallman, T.R., & McCray, R., 1982, Ap. J. Supp. 50, 263.
Kraemer, S., 1986 Ap.J., 307,478
Netzer, H., 1993, ApJ 411, 594.
Netzer, H., & Ferland, G.J., 1984, Publ A.S.P.
Osterbrock, D.E., 1989, *Astrophysics of Gaseous Neublae and Active Galactic Nuclei*, (University Science Books; Mill Valley).
Péquignot, D., 1986, Editor, *Workshop on Model Nebulae*, Publication de l'Observatoire de Paris.
Petitjean, P., Boisson, C., & Péquignot, D., 1990, Ast Ap 240, 433.
Raymond, J.C. 1979, ApJS, 39, 1.
Rees, M., Netzer, H., & Ferland, G., 1989 Ap.J. 347, 640
Rubin, R. H., Dufour, R. J., & Walter, D. K. 1993, ApJ, 413, 242
Rubin, R. H., Simpson, J. P., Haas, M. R., & Erickson, E. F. 1991, ApJ, 374, 564
Seaton, M., 1987, J.Phys.B. 20, 6363.
Shull, P. 1983, ApJ, 275, 611.
Simpson, J. P., Colgan, S. W. J., Rubin, R. H., Erickson, E. F., & Haas, M. R. 1995, ApJ, submitted
Sutherland, R., & Dopita, M., 1993 ApJS, 88, 253
Verner, D., & Yakovlev, D., 1995, Ast Ap. Sup 109, 125.
Viegas, S., & Contini 1994, ApJ 428, 113
Wills, B., Netzer, H., & Wills, 1985, Ap.J. 288, 94.